\documentstyle[psfig]{l-aa} 

\begin{document} 
%=============================================================
%        Additional definitions
%============================================================= 
% 
\def\sco{Nova Sco 1994} 
\def\gro16{GRO~J1655--40} 
\def\ea{\hbox{et~al.}} 
%
%============================================================= 

%================== Title Page =============================== 

\thesaurus{06 
(08.02.1;              % Stars: binaries: close 
08.09.02;              % Stars: individual 
13.25.5)               % X-rays: stars 
} 

\title{The quiescence optical light curve of Nova Scorpii 1994 
(=GRO~J1655--40)\thanks{Based on observations made at the European 
Southern
Observatory, La Silla, Chile}} 

\author{F.~van der Hooft\inst{1}, 
M.H.M.~Heemskerk\inst{1}, 
F.~Alberts\inst{1}, and 
J.~van Paradijs\inst{1,2} } 

\offprints{F. van der Hooft} 

\institute{ Astronomical Institute ``Anton Pannekoek'', University of Amsterdam
and Center for High Energy Astrophysics, Kruislaan 403, NL-1098 SJ Amsterdam,
The Netherlands \and Department of Physics, University of Alabama in
Huntsville, Huntsville, AL 35899, USA } 

\date{Received date; accepted date} 

\maketitle 
\markboth{F. van der Hooft \ea: The quiescence optical light curve of Nova Scorpii 
1994 (=GRO~J1655--40)}{F. van der Hooft \ea: The quiescence optical light curve of 
Nova Scorpii 1994 (=GRO~J1655--40)} 

%================== Abstract =============================== 

\begin{abstract} 

We report on extensive {\it V}, {\it R}, and {\it i} band photometry of the
black-hole candidate \sco\ (\gro16) obtained  during March 1996, when the
source was close to its quiescent pre-outburst optical brightness ({\it
V}=17.3). From our observations  and data taken from the literature we derive a
refined ephemeris for inferior conjunction of the secondary star:  HJD
$244\,9838.4198(52) + 2.62168(14) \times {\rm N}$.  We have modeled the {\it
V}, {\it R}, and {\it i} band light curves in terms of a Roche lobe filling
secondary and an accretion disk around the compact star, the  latter described 
as a flat cylinder with a radial temperature distribution. From the shape of 
the light curve we constrain the binary inclination and mass of the secondary 
star to lie in the ranges $63\fdg7$--$70\fdg7$ and 1.60--3.10 M$_{\odot}$, 
respectively. This limits the mass of the black-hole to the range 6.29--7.60
M$_{\odot}$. The mass range we obtained for the secondary star is supported by 
the results of stellar evolution calculations.

\keywords{{\em (Stars:)} binaries: close, individual: \sco\ (\gro16) -- X-rays:
stars} 

\end{abstract} 

%================== Text and Acknowledgements =============================== 

\section{Introduction} 

Soft X-ray transients (SXTs) are low-mass X-ray binaries consisting of a
neutron star or black-hole primary, which undergo unstable accretion from a
late-type  companion. Disk instabilities (\cite{jvp96}) cause  brief but
violent outbursts, typically lasting weeks to months, during which  their X-ray
luminosity abruptly increases by several orders of magnitude, to 
$\sim10^{37}-5 \times 10^{38}$ erg s$^{-1}$. During such an outburst, the 
optical flux becomes dominated  by the contribution of the accretion disk and
is enhanced by several magnitudes. In the intervals between outbursts, which
typically last years to decades, the X-ray luminosity is extremely low, and the
optical  flux is dominated by the luminosity of the secondary star (see Verbunt
\ea\ 1994, Van Paradijs \& McClintock 1995 and Narayan, McClintock \& Yi 1996
for reviews of the quiescence X-ray and optical properties of SXTs). Therefore,
the properties of the secondary star can be studied during the quiescent
episode. 

The SXT \sco\ (\gro16), discovered on 1994 July 27 with BATSE (20--100 keV) on
board the Compton Gamma Ray Observatory (\cite{zha94}), has been studied
extensively during the past two years in X-rays and at optical and radio
wavelengths (Tingay \ea\ 1995, Bailyn \ea\ 1995a, Harmon \ea\ 1995b, Hjellming 
\& Rupen 1995, Zhang \ea\ 1995, Bailyn \ea\ 1995b, Paciesas \ea\ 1996, Van der 
Hooft \ea\ 1997). CCD
photometry of \sco\ revealed a double-waved modulation of the optical light at
a period of 2.6 days (\cite{bai95b,vdh97}). Strong evidence that the compact
object in \sco\ is a black hole was presented by Bailyn \ea\ (1995b) who
established a spectroscopic period of $2.601\pm0.027$ days and a mass function
$f(M)=3.16 \pm 0.15$~M$_{\odot}$. The secondary was classified as an F3-F6 IV 
type star (\cite{oro97}). 

As an SXT, \sco\ is remarkable, due to the fast recurrence of its successive
X-ray outbursts since the first X-ray detection of the source in 1994. The
X-ray outbursts of \sco\ were initially separated by intervals of about 120
days (\cite{zha95}) and lasted many weeks (Harmon \ea\ 1995a, b; Wilson \ea\
1995). During these periods the optical brightness of \sco\ was between 1 and 2
magnitudes above its quiescent {\it V}=17.3 pre-outburst level (\cite{bai95a}),
which likely reflects the contribution of the accretion disk and X-ray heating
of the secondary star. Most photometric observations of \sco\ were made during
this period. 

However, in the interval 1995 September to 1996 April, no significant X-ray 
flux was detected from  \sco\, indicating a (temporary) longer period of 
quiescence. We present in this paper the results of extensive {\it V}, {\it R}, 
and {\it i} band photometry of \sco\ obtained during 1996 March, and derive its
quiescence light- and color curves. We derive limits on the system parameters
based on $\sim30\,000$ theoretical light- and color curves in the three
photometric bands, calculated for a wide range of parameters of a model
describing a Roche lobe filling secondary star supplemented with a model of an
accretion disk around the compact star. The limits on the mass of the secondary 
star are compared with theoretical evolutionary tracks which we computed for 
both single and binary star evolution.

\section{Observations and Data Reduction} 

\begin{figure*} 
\psfig{figure=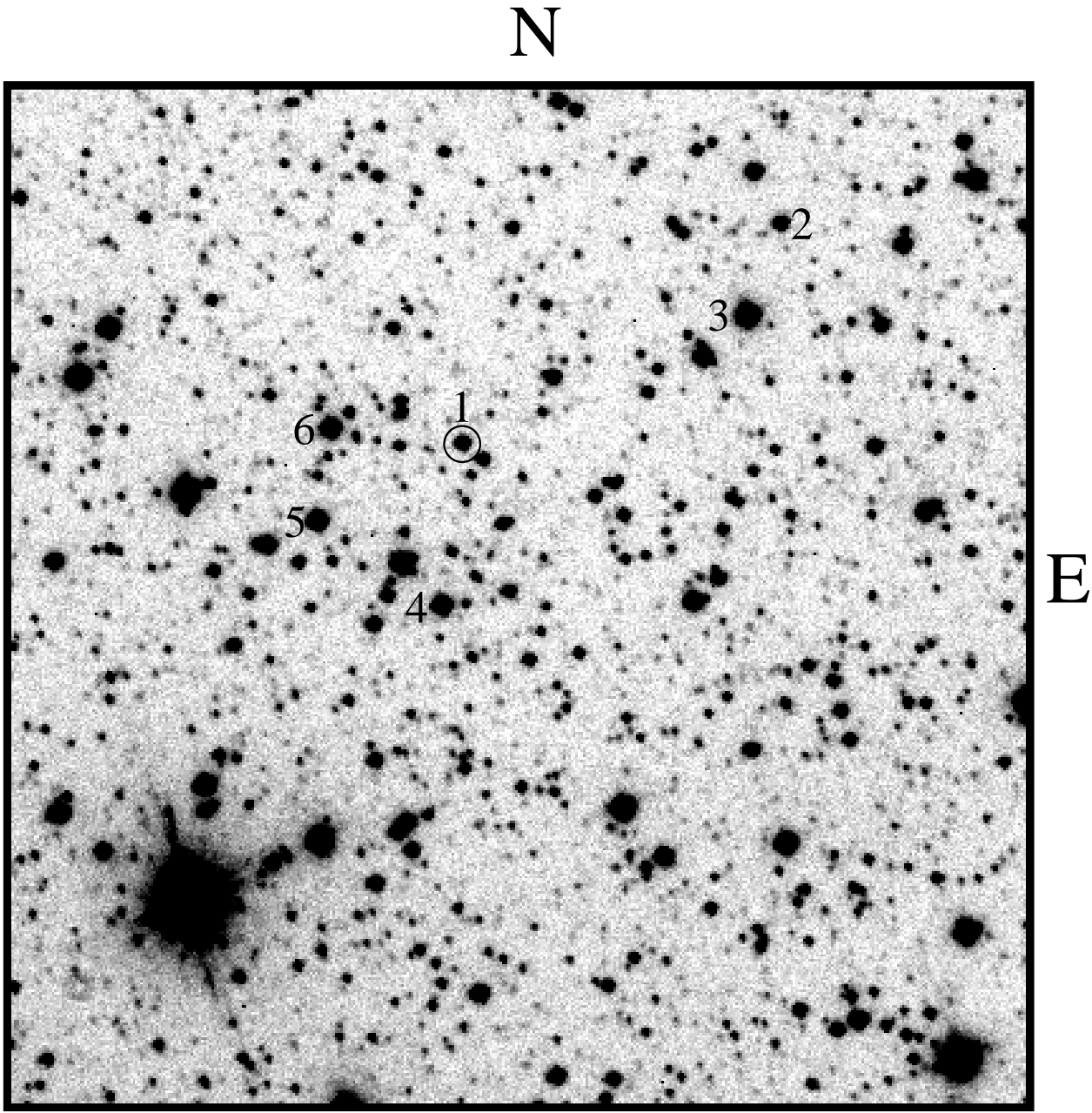} 
\caption[]{$3\arcmin \times
3\arcmin$ {\it R}-band find chart of the field of \sco, taken on 1996 March 5.
The exposure time was 300 seconds; North is at the top, East to the right.
\sco\ is indicated by a circle and is numbered 1, the five comparison stars are
numbered 2-6. 
} 
\label{fig1} 
\end{figure*}

We observed \sco\ (\gro16) during 28 consecutive nights starting 1996 March 5
with the 0.91m Dutch telescope at the European Southern Observatory (ESO) in
Chile. The telescope was equipped with a CCD camera and the observations were
made using standard {\it V} and {\it R} filters and a Gunn {\it i} filter. 
Integration times were 10 minutes for the {\it V} filter and 5 minutes for the 
{\it R} and {\it i} filters. We obtained 67 images of \sco\ in the {\it V}
band, and 69 in the {\it R} and {\it i} bands each. All images were corrected
for the bias, and flat fielded in the standard way. A $3\arcmin \times
3\arcmin$ {\it R}-band find chart of the field of \sco\ (taken on 1996 March
5), in which \sco\ and five nearby comparison stars are indicated, is shown in
Fig.~\ref{fig1}. 

We applied aperture photometry to \sco\ and five nearby comparison stars using 
MIDAS and additionally written software operating in the MIDAS environment. The 
five comparison stars were checked for variability over the entire
data set. From the disperion of the relative magnitudes of each comparison
star over the entire observing period ($\leq 0.008$ mag) we conclude that the 
brightness of these stars is constant. The accuracy of a single measurement of
the differential magnitude of \sco\ (relative to the average of the five
comparison stars), taking into account Poisson noise only, is typically 0.006
mag in each of the three bands. 

Photometric calibration of the data was performed using images of two Landolt
standard  fields (\cite{lan92}), Rubin 149 (8 stars) and PG~1323 (4 stars),
during nights of good photometric quality. Observations of different standard
fields within one night showed that the accuracy of the photometric
calibration was better then 0.05 mag. 

\section{Results} 

\subsection{Quiescence light and color curves of \sco} 

During the period of our observations the average {\it V} magnitude of \sco\ 
was 17.25 and the light curve showed a modulation with two equal maxima and 
unequal minima, whose depths are 0.31 and 0.48 magnitudes. The average {\it V}
magnitude of \sco\ is  consistent with the pre-outburst brightness of {\it
V}=17.3 (\cite{bai95b}), which indicates that the source was in quiescence
during the period of our observations. During two successive pointings on
March 8, separated by $\sim5$ hours, we detected a very low optical
brightness of \sco\ in each of the three filters (for the {\it V}-band we then
found {\it V}=17.48 and {\it V}=17.53, respectively). 

There is remarkably little variability superposed on the quiescence light curve
of \sco\ in comparison to several other quiescent SXTs, which show 
cycle-to-cycle variations of the order of a few percent
(McClintock \& Remillard 1986, Remillard \ea\ 1992, Martin \ea\ 1995, Chevalier 
\& Ilovaisky 1996, Shahbaz \ea\ 1996). Haswell (1996) has recently suggested
that such brightness variations may be similar to those seen in superhumps
during superoutbursts of SU UMa type cataclysmic variables. 

The data in the three bands were independently searched for periodicities using
a Lomb-Scargle periodogram (\cite{pre89}). The periodograms show five strong
peaks, corresponding to half the orbital period and its  one-day aliases, and
associated with each peak the characteristic pattern resulting from the data
sampling. In each band we found a distinct maximum at a fundamental period of
1.3123 days. Since the light curve of \sco\ features two minima with unequal
depth, we folded the data on twice this period, using a trial zero point
(chosen in the center of the data set). The time of inferior conjunction of the 
secondary, $T_{\rm inf}$, was derived from the folded light curve, by 
fitting a parabola to the shallow minimum, which yielded a phase offset to be
applied to the trial zero point. The shallow minimum occurs close to the epoch
of inferior conjunction of the secondary star as predicted by the ephemerides
of \sco\ reported in previous investigations (see e.g. Bailyn \ea\ 1995b, Van
der Hooft \ea\ 1997). 

\begin{figure*} 
\psfig{figure=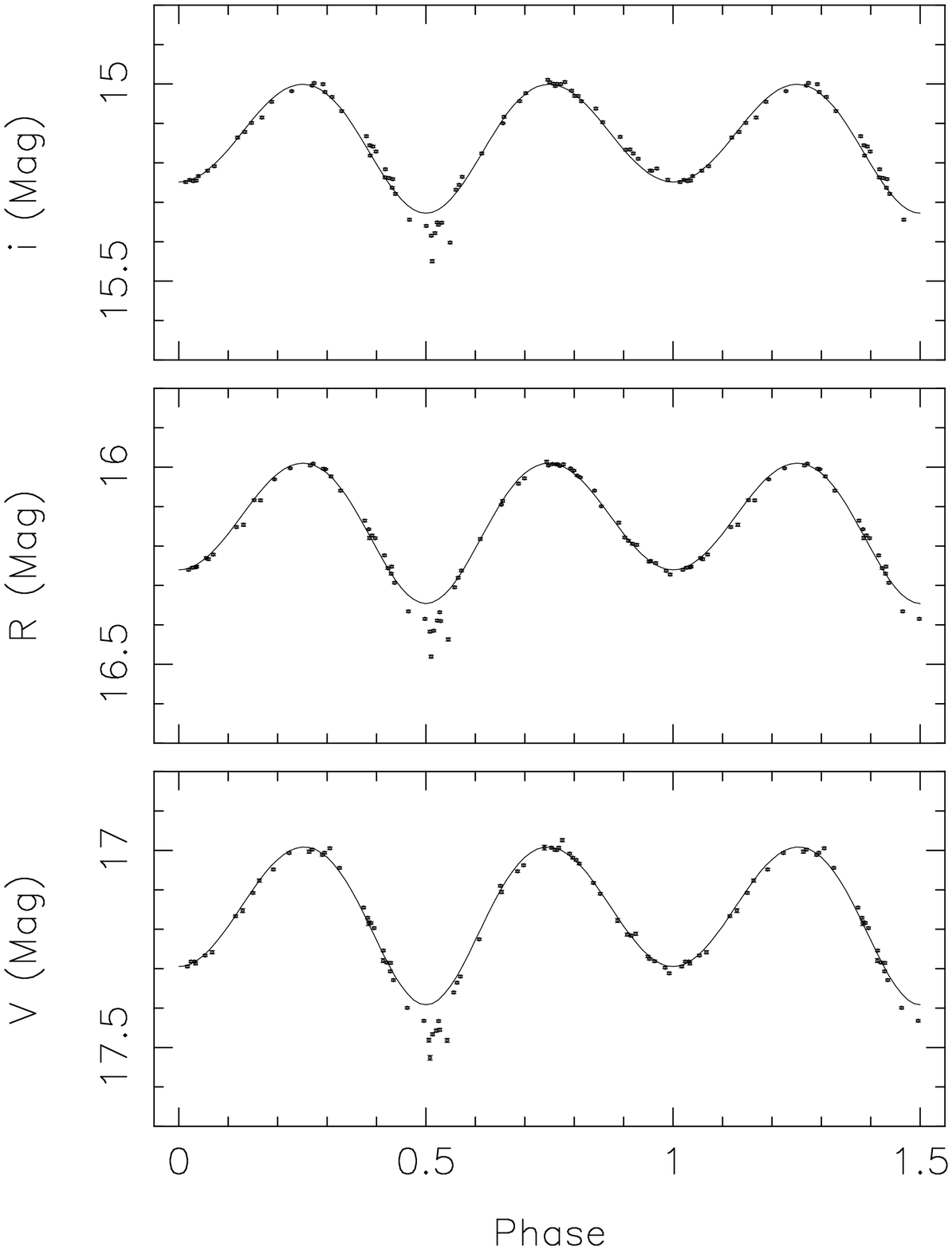,height=22cm} 
\caption[]{{\it V}, {\it R}
and {\it i} light curves of \sco\ folded at the photometric ephemeris $T_{\rm
inf}({\rm HJD})=244\,9838.4198(52) + 2.62168(14) \times {\rm N}$. The curves
are repeated over 1.5 cycles for clarity. Phase 0.0 corresponds to inferior
conjunction of the secondary with respect to the compact star. The drawn lines
represent a theoretical light curve computed for a binary inclination and
secondary mass of ($67\fdg25$, 1.60 M$_\odot$).
} 
\label{fig2} 
\end{figure*}

The photometric period of 2.6246 days of \sco\ is slightly longer than previous
measurements of the period based on both photometric and spectroscopic
observations (see e.g. Bailyn \ea\ 1995b, Van der Hooft \ea\ 1997). In order to
improve the accuracy of the orbital period we combined 9 photometric
measurements of $T_{\rm inf}$ from our 1995 and 1996 data with photometric
measurements by Bailyn \ea\ (1995b) and Orosz \& Bailyn (1997), and two
spectroscopic measurements by Casares (1996) and Orosz \& Bailyn (1997). From a
least-squares fit of this set of 13 arrival times to their respective cycle
numbers (spanning 140 cycles) we derived the following refined ephemeris for
the times of inferior conjunction of the secondary star: HJD 
$244\,9838.4198(52) + 2.62168(14) \times {\rm N}$, with $\chi^{2}_{\nu}$=2.3
for 11 degrees of freedom. In Fig.~\ref{fig2} we present the {\it V}, {\it R}
and {\it i} light curves of \sco\ folded at this ephemeris. The drawn lines 
represent light curves which were computed using a theoretical model of the 
secondary star which will be described in detail in Sect.~3.2.

\begin{figure*} 
\psfig{figure=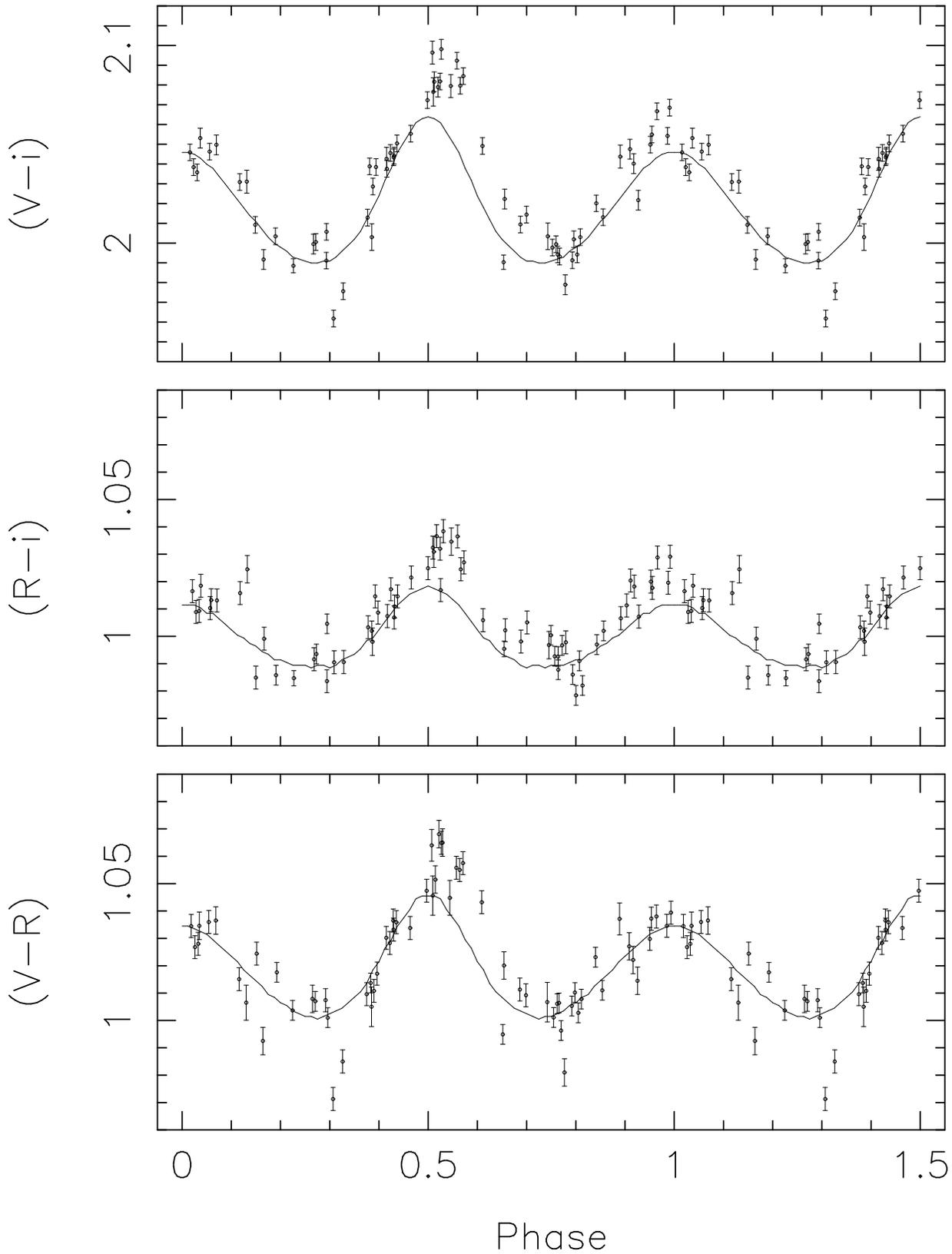,height=22cm} 
\caption[]{{\it (V--R)},
{\it (R--i)} and {\it (V--i)} color curves folded at the photometric ephemeris
$T_{\rm inf}({\rm HJD})=244\,9838.4198(52) + 2.62168(14) \times {\rm N}$. The
curves are repeated over 1.5 cycles for clarity. Phase 0.0 corresponds to
inferior conjunction of the secondary with respect to the compact star. The
drawn lines represent a theoretical color curve computed for a binary
inclination and secondary mass of ($67\fdg25$, 1.60 M$_\odot$).
} 
\label{fig3}
\end{figure*}

We calculated the {\it (V--R)}, {\it (R--i)} and {\it (V--i)} color curves 
using consecutive {\it V}, {\it R} and {\it i} measurements taken close in time 
($\la 20$ min). These color indices, folded at the above ephemeris, are
displayed in Fig.~\ref{fig3}, together with theoretical color curves (see
Sect. 3.2). They  show a clear modulation with the photometric period. The
peak-to-peak amplitudes in the {\it (V--R)} and {\it (R--i)} color curves are
$\sim0.06$  and $\sim0.05$ mag, respectively. The color curves have two
unequal maxima and two equal  minima. The maxima of the color curves occur near
phase 0.0 and 0.5 (the largest occurs near phase 0.5), indicating that \sco\
is reddest near conjunctions; the system is bluest near quadratures (phases
0.25 and 0.75). 

\subsection{Modelling the light curves: ellipsoidal model} 

To construct theoretical light curves we calculated the flux of the secondary
emitted into the direction of the observer as a function of orbital phase, by
numerical integration of the contributions from a large number of small
elements on the secondary's surface. We used a numerical model, discussed in
detail by Tjemkes \ea\ (1986), based on a description
of the tidal and rotational distortion of the secondary in terms  of an
equipotential surface. The non-uniform surface brightness distribution across
the secondary is described according to Von Zeipel's theorem. We adopted 0.25
as the value of the gravity darkening coefficient $\beta$, valid for stars
with a radiative envelope (\cite{zei24}). For each surface element we used a
black-body approximation (at the local effective temperature) for the
perpendicularly emerging specific intensity. We applied a standard linear limb
darkening law, with limb darkening  coefficients as a function of temperature
taken from Al-Naimy (1978). For the theoretical spectra calculated by Kurucz
(1979) one finds that over the effective temperature interval 5500 to 8000 K
(covering virtually all surface elements on the secondary) the average
deviation of the gradient $\partial \log F_{\it V}/\partial T$ with respect to
the blackbody approximation is $\sim6$~\% (similar results were obtained for
the dependence of $F_{\it R}$ and $F_{\it i}$ on $T$ [$\sim3-4~\%$]). To
estimate the effect of our blackbody approximation on our results we have made
trial calculations in which the Planck functions were multiplied by an
arbitrary function of $T$ which accounts for the difference with the Kurucz
spectra. We find that the depth of the deepest minimum changes by $\la 2$~\%
({\it V}-band) and $\la 1$~\% ({\it R} and {\it i}-band) with respect to the 
blackbody approximation, while the depth of the shallow minimum changes by 
$\la 4$~\% ({\it V}-band) and $\la 2$~\% ({\it R} and {\it i}-band). These 
differences are small enough that their effect on the solutions in the 
$(i, M_2)$ plane (see below) can be ignored compared to the effects of 
uncertainty in the system parameters. 

Initially, the possible
contribution of an accretion disk to the emergent flux was not included in our
modelling, nor were mutual eclipses and the effect of X-ray heating on the
secondary. In our calculations we used the secondary mass $M_2$ and the orbital
inclination $i$ as free parameters. The observed mass function
$f(M)=3.16\pm0.15$~M$_{\odot}$ (\cite{bai95b}) then fixes the primary mass
$M_1$ and the mass ratio $q=M_1/M_2$. Together with the bolometric luminosity
$L_{\rm opt}$ of the secondary, the temperature distribution across the
latter's surface, and the light curve, are determined. 

We calculated a first set of 81 light curves (and their corresponding color
curves) for $i$ between $50\degr-90\degr$ (steps of $5\degr$), and secondary
masses in the range $0.5 \leq M_2 \leq 4.5$ M$_{\odot}$ (steps of
0.5~M$_{\odot}$). We adopted a bolometric luminosity for \sco\ of {\it L}$_{\rm
opt}=41$ L$_{\odot}$, as determined from the mean {\it V} brightness 
$({\it V}=17.25$, see Fig.~\ref{fig2}), a distance to the system of 3.2 kpc
(\cite{hje95}), a bolometric correction ${\it B.C.}=-0.03$ mag (F5 V type 
star; Popper 1980) and a color excess {\it E(B--V)}=1.3 mag (\cite{hor96}). 
Each of
the theoretical light curves were given an offset such that the data observed
at the shallow minimum (phase 0.0) was matched. We found that none of these
light curves could reproduce the sharp and deep minimum at phase 0.5,
indicating that our model requires additional structure.  Near this phase the
compact object is in front of the secondary star, and the light curve shows
significant scatter whereas it is smooth during the remaining part of the
orbital cycle. Both the sharpness of the minimum and the  increased scatter
near phase 0.5 likely result from the presence of an accretion  disk around the
compact star, which is obscuring (part of) the light of the  secondary at this
phase. There is no evidence for a disturbance of the light curve near phase
0.0; we therefore, in this first phase of our analysis, computed a $\chi_\nu^2$
for each set of parameters, using the light curves in the three photometric
bands simultaneously from which the data obtained between photometric phase
0.35 and  0.65 were excluded. 

We found that the only acceptable solutions are for $i$ in the range 
$60\degr-75\degr$, for any of the secondary masses between 0.5--4.5 
M$_\odot$. We made a new set of theoretical light (and color) curves, using a 
denser grid of $i$ and $M_2$ values: $61\degr \leq i \leq  74\degr$ 
($0\fdg25$ step size) and $0.2 \leq {\it M}_2 \leq 5.0$ M$_\odot$ 
(0.1 M$_\odot$ step size), respectively. 

\begin{figure*} 
\psfig{figure=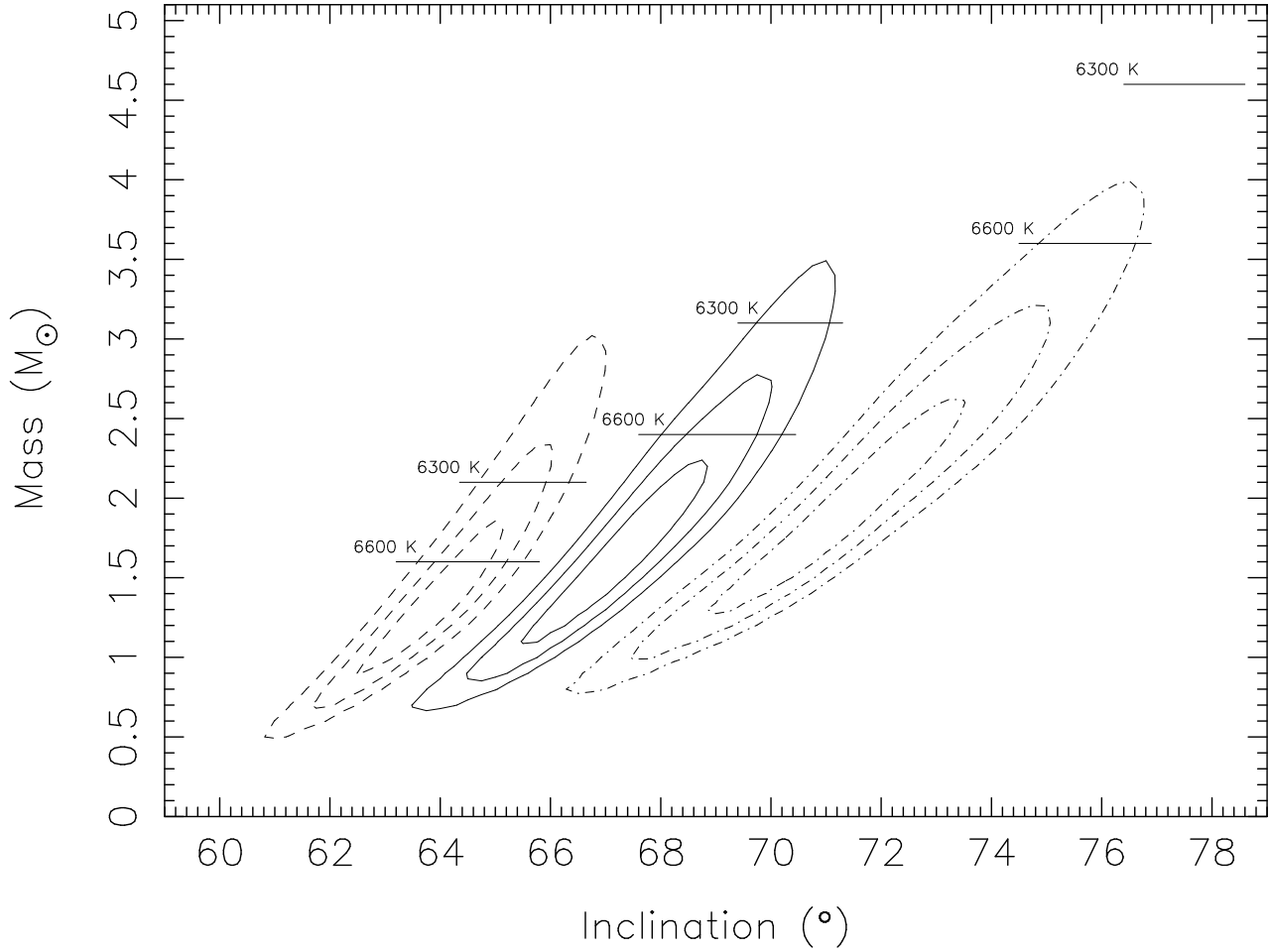,width=17cm,angle=-90} 
\caption[]{
$1, 2, 3\sigma$ confidence contours in the $i$ versus $M_2$ plane for intrinsic
luminosities of 31 (dashed), 41 (solid), and 54 L$_\odot$ (dashed-dotted). The
average effective temperatures from the secondary, calculated from the
intrinsic luminosity and the effective radius of the secondary, are indicated
for 6300 and 6600 K.
} 
\label{fig4} 
\end{figure*}

From the fits of the theoretical to the observed light curves we find that the
binary inclination and mass of the secondary star can be well constrained. The
$\chi_\nu^2$ contours in the $i$ versus $M_2$ plane trace a narrow,
ellipse-shaped region. Such a correlation between $i$ and $M_2$  is expected as
the amplitude of the brightness modulation will decrease with both smaller
inclination angles and larger mass of the secondary  star (the distortion of
the secondary, and therefore the temperature  contrast across the stellar
surface decreases as $M_2$ increases). A minimal $\chi_\nu^2$ of 2.54 (130
d.o.f.) was found at ($67\fdg25$, 1.60 M$_\odot$), which indicates that minor
variations are present in the light curve in the phase interval 0.65--1.35. The
1, 2 and $3\sigma$ confidence contours in the $(i, M_2)$ plane are shown in
Fig.~\ref{fig4}, which were computed by scaling the required $\Delta\chi^2$ 
by the minimum value of $\chi_{\nu}^2$. 

%
% $1\sigma$ confidence level for two parameters; $\delta chi^2=2.30$ required.
% $2\sigma$ confidence level for two parameters; $\delta chi^2=6.17$ required.
% $3\sigma$ confidence level for two parameters; $\delta chi^2=11.8$ required. 

The assumed value of the bolometric luminosity of the secondary star of \sco\ 
(41 L$_{\odot}$) is  sensitive to errors in both the distance to the
source and its color excess. The latter is probably the most important source
of  uncertainty as the distance is well determined at $3.2 \pm 0.2$ kpc
(\cite{hje95}) based on a kinematic model of the radio jets; other distance
estimates are consistent with this value  [3.5 kpc (\cite{mck94}); 3--5 kpc
(\cite{tin95}) both based on H {\sc i}  absorption; $\sim3$ kpc (\cite{bai95a})
based on interstellar absorption; $\sim3$ kpc (\cite{gre95}) based on a dust
scattering halo]. A minimal value of the color excess was reported by Bailyn
\ea\ (1995a) [{\it E(B--V)}=1.15 based on the equivalent width of the Na
D-lines in optical spectra taken in 1994 August], while a largest {\it E(B--V)}
of 1.5 was reported by Horne \ea\ (1996) [based on a power-law spectral fit to
{\it UBVRI} photometry taken during 1996 May]; deep 220-nm absorption in a HST
spectrum of \sco\ taken in 1996 May suggested {\it E(B--V)} of 1.3
(\cite{hor96}). Therefore, we assumed {\it E(B--V)} to lie in the range
1.2--1.4, with the corresponding value of {\it L}$_{\rm opt}$ in the range
31--54 L$_\odot$.  We therefore also calculated theoretical light- and color
curves for secondary luminosities  of 31 (for which we took $59\degr \leq i
\leq 74\degr$), and 54 L$_\odot$ (with $62\degr \leq i \leq 78\degr$); the
steps in $i$ were $0\fdg25$. In all calculations we covered the range  $0.2
\leq M_2 \leq 5.0$  M$_\odot$, respectively, with steps of 0.10 M$_\odot$. 

From these theoretical light curves we derived ellipse-shaped $\chi_\nu^2$
contours in the $i$ versus $M_2$ plane similar to those derived for the curves
calculated for $L_{\rm opt}=41$~L$_\odot$. However, the location of the
contours in the $(i, M_2)$ plane  depends on $L_{\rm opt}$. For 31 L$_\odot$ 
the $\chi_\nu^2$ contours shift towards lower binary inclinations, whereas they
 shift to larger inclination for 54 L$_\odot$. Minimal $\chi_\nu^2$  values
(for 130 d.o.f.) were found at  ($63\fdg75$, 1.30 M$_\odot$)
[$\chi_\nu^2=2.60$; 31 L$_\odot$], and  ($71\fdg25$, 1.90 M$_\odot$)
[$\chi_\nu^2=2.52$; 54 L$_\odot$]. The 1, 2 and $3\sigma$ confidence contours
were computed similar as in the case of 41 L$_\odot$, and are displayed in
Fig.~\ref{fig4}.

\subsection{Inclusion of an accretion disk to the model} 

The model described in Sect. 3.2 quite well describes the data obtained at
inferior conjunction of the secondary star, but not those obtained near superior
conjunction, when we observe less light than the model predicts. A natural
explanation for the flux deficit near  phase 0.5 is that an accretion disk
around the compact star obscures part of the secondary at certain viewing
angles. In the second stage of our modelling efforts we included an accretion
disk in the model, described as a flat cylinder with a radial temperature
distribution. The vertical outer edge of the disk is assumed not to emit
radiation. By including such an accretion disk the number of parameters in the
model is increased by three: the radius of the disk as a fraction of the
effective Roche lobe, $\alpha$, the flaring angle $\gamma$ of the disk, and 
the temperature at the outer edge of the disk, $T_{\rm edge}$. In view of the
very low X-ray luminosity of \sco\ during the period of our observations, we
assumed that the radial temperature distribution in the disk follows an
$r^{-3/4}$ dependence (Pringle 1981). 

We computed $\sim25\,000$ theoretical light and color curves for intrinsic 
luminosities of 31, 41 and 54 L$_\odot$, a fractional disk radius between  0.6
and 1.0 (step size of 0.1), outer disk temperatures of 100 or 1000 K, and a
flaring angle of the disk of either $2\degr$ or $10\degr$. We used step sizes
of $0\fdg25$ and 0.1 M$_\odot$ for $i$ and $M_2$, respectively. We performed
several tests (using a less dense grid for $i$, and $M_2$) in which temperatures
at the outer edge of the disk were selected between 100 and 6000 K. Light
curves computed with $T_{\rm edge}>1000$ K do not resemble the data well: for 
such outer edge temperatures the accretion disk contributes too much to the 
total flux of the system, which decreases the amplitude of the ellipsoidal 
modulation and changes the relative depth of the minima in the light curve. 
At superior conjunction of the secondary star, the contribution of the disk to 
the total flux is $\leq 4~\%$ for {\it T}$_{\rm edge}=1000$ K, and is 
entirely negligible for lower values of {\it T}$_{\rm edge}$. 
Thus, the main effect 
of the disk is to occult a fraction of the secondary star near phase 0.5. 

In calculating $\chi_\nu^2$ we did not include the six deviating points (two in
each of the three passbands) obtained during subsequent pointings on March 8;
they must be caused by variations in the disk structure, which cannot be
incorporated into our model. 

\begin{figure*} 
\hbox{ 
\psfig{figure=fig5ul.ps,angle=-90,width=9cm}
\psfig{figure=fig5ur.ps,angle=-90,width=9cm} 
} 
\hbox{
\psfig{figure=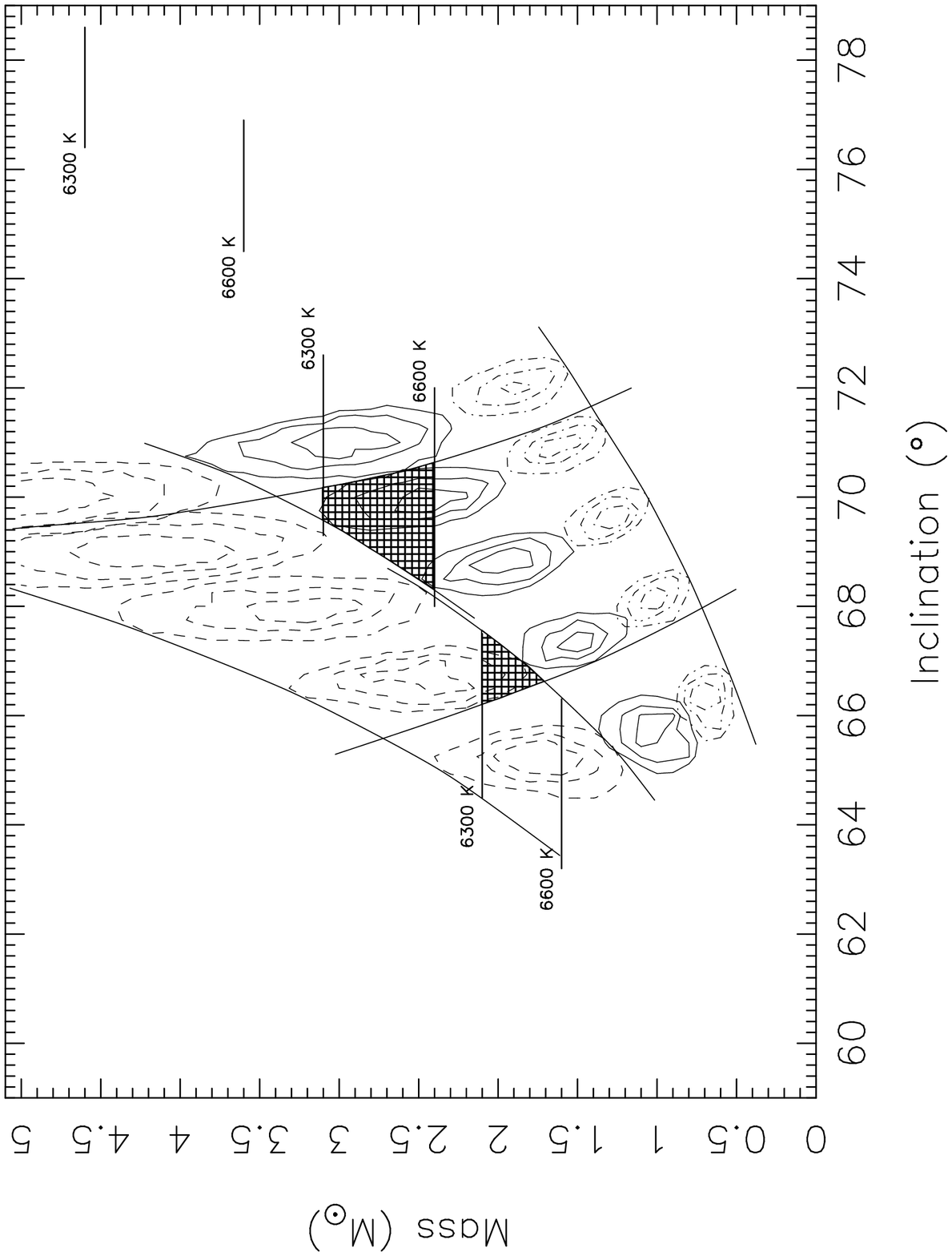,angle=-90,width=9cm}
\psfig{figure=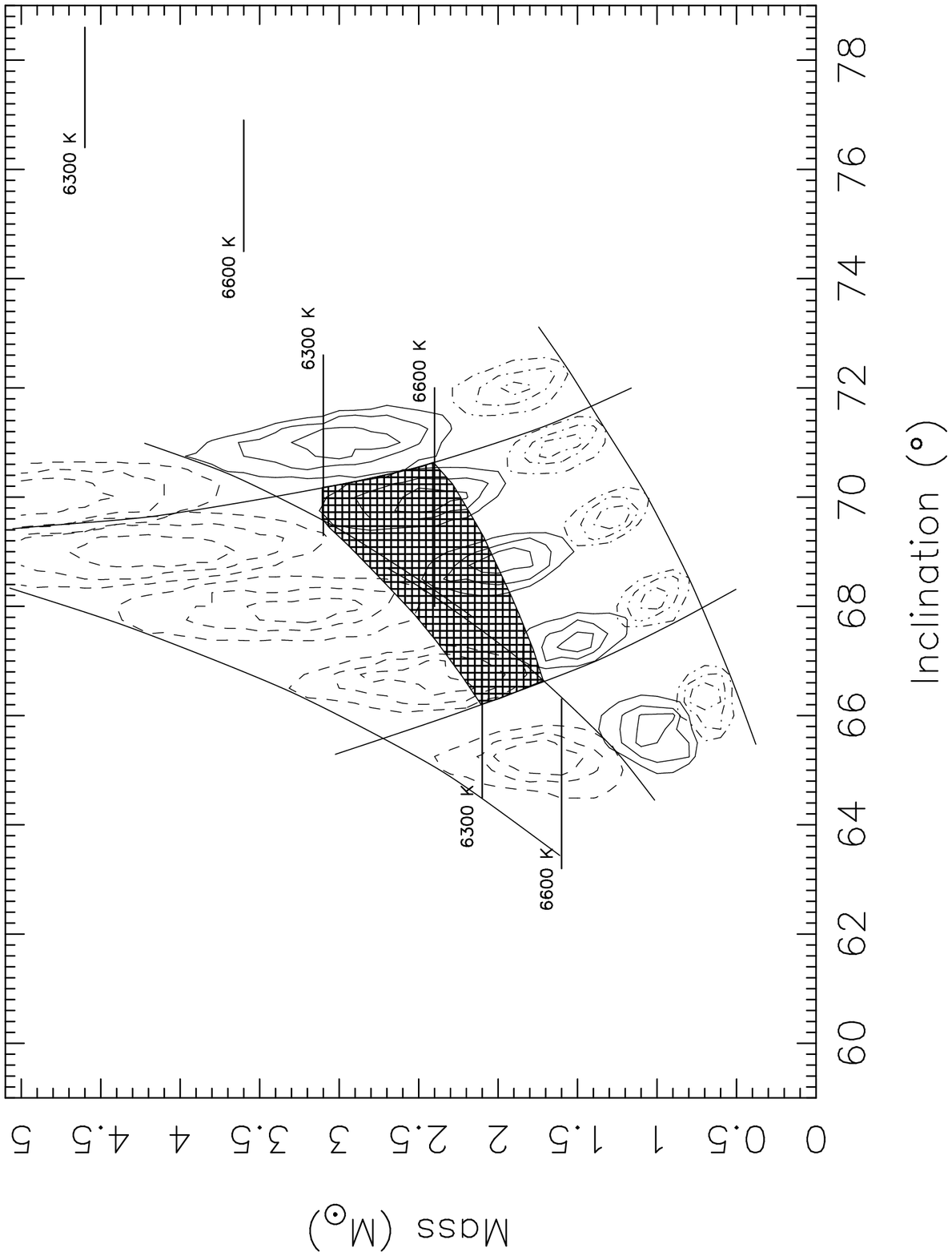,angle=-90,width=9cm} 
} 
\caption[]{ $1, 2, 3\sigma$
confidence contours in the $i$ versus $M_2$ plane for intrinsic luminosities of
41 [top-left panel], and 31 (dashed), 41 (solid), and 54 L$_\odot$
(dashed-dotted) [other panels], for accretion disk models with outer edge
temperature {\it T$_{\rm edge}$}=100 K, flaring angle $\gamma=2\degr$, and
fractional disk radii between 0.6 and 1.0. Average effective temperatures of
the secondary star are indicated in the top-right panel. In the bottom-left
panel, the solutions are constrained by the smooth curves tangent to the
$\alpha=0.7, 0.9$, and 31, 41 and 54 L$_\odot$ confidence contours. The
solutions in the $i$ versus $M_2$ plane which are constrained by these limits
are indicated by the hatched area in the bottom-left panel. Finally, the total
collection of solutions for this ({\it T$_{\rm edge}$}, $\gamma$) is given in
the bottom-right panel. 
} 
\label{fig5} 
\end{figure*}

For each set of parameters, the 1, 2 and $3\sigma$ confidence contours in the
$(i, M_2)$ diagram were computed as before. The confidence contours
corresponding to models having an accretion disk with a flaring angle 
$\gamma=2\degr$, and an outer-edge temperature  ${\it T}_{\rm edge}=100$ K, 
are shown
in Fig.~\ref{fig5}. The top-left panel of this figure displays the
confidence contours for $L_{\rm opt}=41$~L$_\odot$ and fractional disk radii
between 0.6 and 1.0. These contours  trace narrow ($\sim1\fdg5$ wide) ellipses
in the $(i, M_2)$ plane. Since the light curve of \sco\ is dominated by
ellipsoidal variations (see Sect. 3.2), the `islands' derived for the 
different fractional disk radii are located in the same general area of the
$(i, M_2)$ diagram as the contours which were derived from the pure ellipsoidal 
model (see Fig.~\ref{fig4}). As expected, for increasing values of
$\alpha$ the confidence contours move toward smaller values of $i$. 

In the top-right panel of Fig.~\ref{fig5}, the confidence contours for
intrinsic luminosities of 31 (dashed), 41 (solid) and 54 L$_\odot$
(dashed-dotted) are  displayed, with the other parameters unchanged. From this
panel it follows that for a given fractional disk size, and intrinsic
luminosities between 31 and 54 L$_\odot$ the solutions move along a rather
narrow track in the $(i, M_2)$ diagram. Solutions for larger secondary masses
are found at smaller inclination angles, since an accretion disk of a given
relative size can obscure the light of the secondary star at smaller values of
$i$ as $M_2$, and therefore, the size of the secondary, increases. In the
remainder of our analysis we have assumed that the fractional disk radius is in
the range 0.7 to 0.9 (see e.g. Paczy\'{n}ski 1977, and Frank \ea\ 1992). 

We can further constrain the system parameters by noting that the spectral type
of the secondary limits the range of its effective temperature (averaged over
the secondary surface), and therefore of the secondary radius. Since the
secondary is assumed to fill its Roche lobe (this fixes its average density),
for a given value of $L_{\rm opt}$ this requirement limits the position of the
system in the $(i, M_2)$ diagram. 
The effective temperature decreases with both smaller intrinsic luminosity  and
larger mass of the secondary star (for a Roche lobe filling secondary the
average density depends only on orbital period). The secondary of \sco\ has
been  classified as an F3-F6 type star (\cite{oro97}). According to Popper
(1980), effective temperatures of such stars are in the range 6330--6620 K. 
The $M_2$ ranges corresponding to this range in average effective temperatures
have been indicated for each intrinsic luminosity individually in the top-right 
panel of Fig.~\ref{fig5}. In the bottom-left panel of Fig.~\ref{fig5} this 
constraint is combined with those obtained from the fitting of the light 
curves. 

For $L_{\rm opt}=54~{\rm L}_\odot$, average effective temperatures of 
$\leq 6600$
K are obtained for secondary masses of 3.6 M$_\odot$ and higher. This limiting
mass is much larger than the maximum mass of the secondary allowed by the disk
models for this luminosity. Therefore, taking the mass limit imposed by the
average effective temperature of the secondary star into account, we can
exclude an intrinsic luminosity of 54 L$_\odot$ for \sco\ (see top-right 
panel of Fig.~\ref{fig5}). For the other two
intrinsic luminosities, the average effective temperature limits the secondary
mass to the range 1.6--2.1 M$_\odot$ (31 L$_\odot$) and 2.4--3.1 M$_\odot$ (41
L$_\odot$), well inside the collection of solutions for these disk models. The
combined constraints for these two values of $L_{\rm opt}$ have been indicated
by hatched polygons in the lower-left panel of Fig.~\ref{fig5}. 

Given the smooth variation of the location of the confidence contours in the
($i, M_2$) plane in response to changes in the fractional disk radius or
bolometric luminosity, we connected both hatched areas by smooth curves,
thereby producing a final constraint on the system in the $(i, M_2)$ plane,
which is shown in the lower-right panel of Fig.~\ref{fig5}. Note that the  
final constraint shown in Fig.~\ref{fig5} is derived for the case 
$T_{\rm edge}=100$ K and $\gamma=2\degr$ only. 

In Fig.~\ref{fig6} we have collected the results of similar analyses for four
combinations  of outer-edge temperature and flaring angle of the disk: (100 K;
$2\degr$) [top-left panel], (1000 K; $2\degr$) [top-right panel], (100 K;
$10\degr$) [bottom-left panel], and (1000 K; $10\degr$) [bottom-right panel].
This figure shows that the solutions corresponding to models for ${\it T}_{\rm
edge}=1000$ K (right panels) are not shifted much  with respect to those for
100 K (left panels). The models with a flaring angle of $\gamma=10\degr$
(bottom panels) are slightly shifted toward lower binary inclination with
respect to the models for $\gamma=2\degr$ (top panels). 

In Fig.~\ref{fig7} and \ref{fig8} theoretical light- and color 
curves are shown, computed for the parameters $i=68\fdg75$, 
$M_2=2.10~{\rm M}_\odot$, 41 L$_\odot$, $\alpha=0.8$, 
${\it T}_{\rm edge}=100$ K, $\gamma=2\degr$, typical for the collection of 
solutions we obtained. 

\section{Discussion} 
\subsection{Quiescence light curves} 

The average {\it V} magnitude of \sco\ during our March 1996 observations  was
17.25, consistent with the pre-outburst brightness of {\it V}=17.3
(\cite{bai95a}), which indicates that the source was then in quiescence. This
is confirmed by X-ray observations of \sco\ made with ASCA and ROSAT during
March 1996, which showed that its X-ray luminosity then was $2 \times 10^{32}$
erg/s (Robinson 1997). No significant X-ray flux of \sco\ was detected
with RXTE during the period of our observations (3$\sigma$ upper limit of 12
mCrab in the 2--12 keV band); X-rays were first detected from \sco\ with RXTE on
April 23, 1996 (Remillard \ea\ 1996). 

Also, the smoothness of the {\it V}, {\it R} and {\it i} light  curves
(spanning 11 consecutive cycles of \sco) is in contrast to earlier reports of
significant cycle-to-cycle variations in the \sco\ light curve  during periods
of known X-ray activity (\cite{bai95b,vdh97}). The relative depth of the two
minima in the light curves of \sco\ is consistent with a quiescence optical
light curve caused by ellipsoidal variations, the deepest minimum occuring at
superior conjunction of the  secondary star as expected.  The brightening in 
X-rays in late April 1996 was followed by increased activity at optical, and
radio wavelengths (Horne \ea\ 1996, Hunstead \& Campbell-Wilson 1996, Hjellming 
\& Rupen 1996).

We determined a revised ephemeris for inferior conjunction of the secondary 
star with respect to the black hole by combining our measurements of  $T_{\rm
inf}$ with those published in other investigations, spanning a total of 140
orbits of \sco. This period differs by $2.7\sigma$ from the period
we derived previously (\cite{vdh97}), and $0.8\sigma$ from the spectroscopic
period derived by Orosz \& Bailyn (1997). 

The colours of \sco\ show a double-waved orbital modulation, with amplitudes of
$\sim0.06$ in {\it (V--R)}, $\sim0.05$ in {\it (R--i)} and $\sim0.11$ in 
{\it (V--i)}. The system is reddest near the conjunctions, bluest at the 
quadratures. 
This colour variation reflects the varying temperature across the surface of
the secondary; its pronounced appearance is the result of the wavelength and
temperature dependence of the limb darkening, which gives relatively low weight
to emission from near the apparent horizon of the secondary. 

The light curve of \sco\ can be understood in terms of a model of a distorted
secondary star in combination with an accretion disk. The disk eclipses part
of the secondary near phase 0.5, but does not give a significant contribution 
to the total luminosity of the system. The largest variability in the light 
curve of \sco\ occurs close to phase 0.5, and is most likely caused by 
variability in the structure of the outer disk, causing variability in the 
fraction of the secondary that is being eclipsed. 

\subsection{Mass determination} 
\begin{figure*} 
\hbox{ 
\psfig{figure=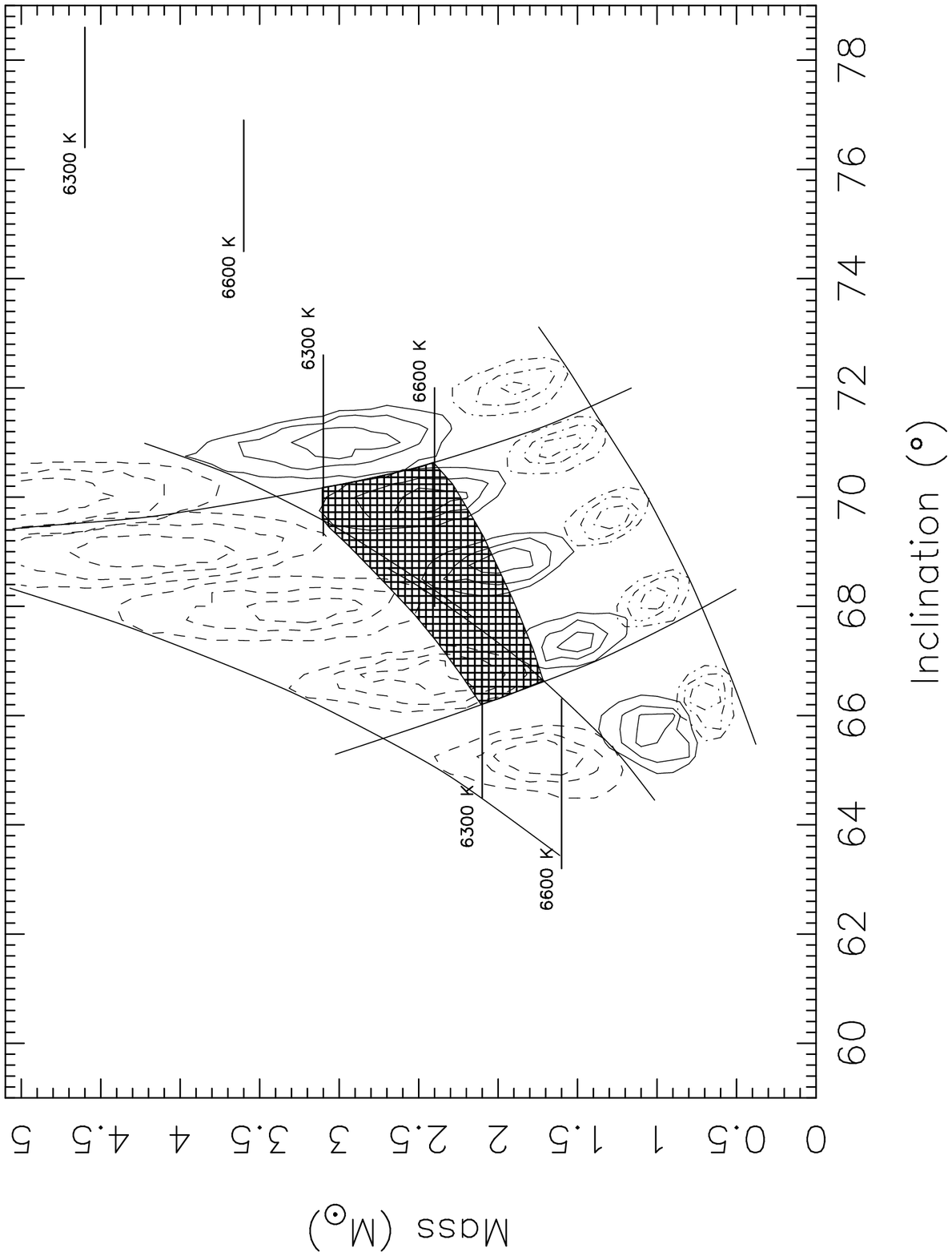,angle=-90,width=9cm}
\psfig{figure=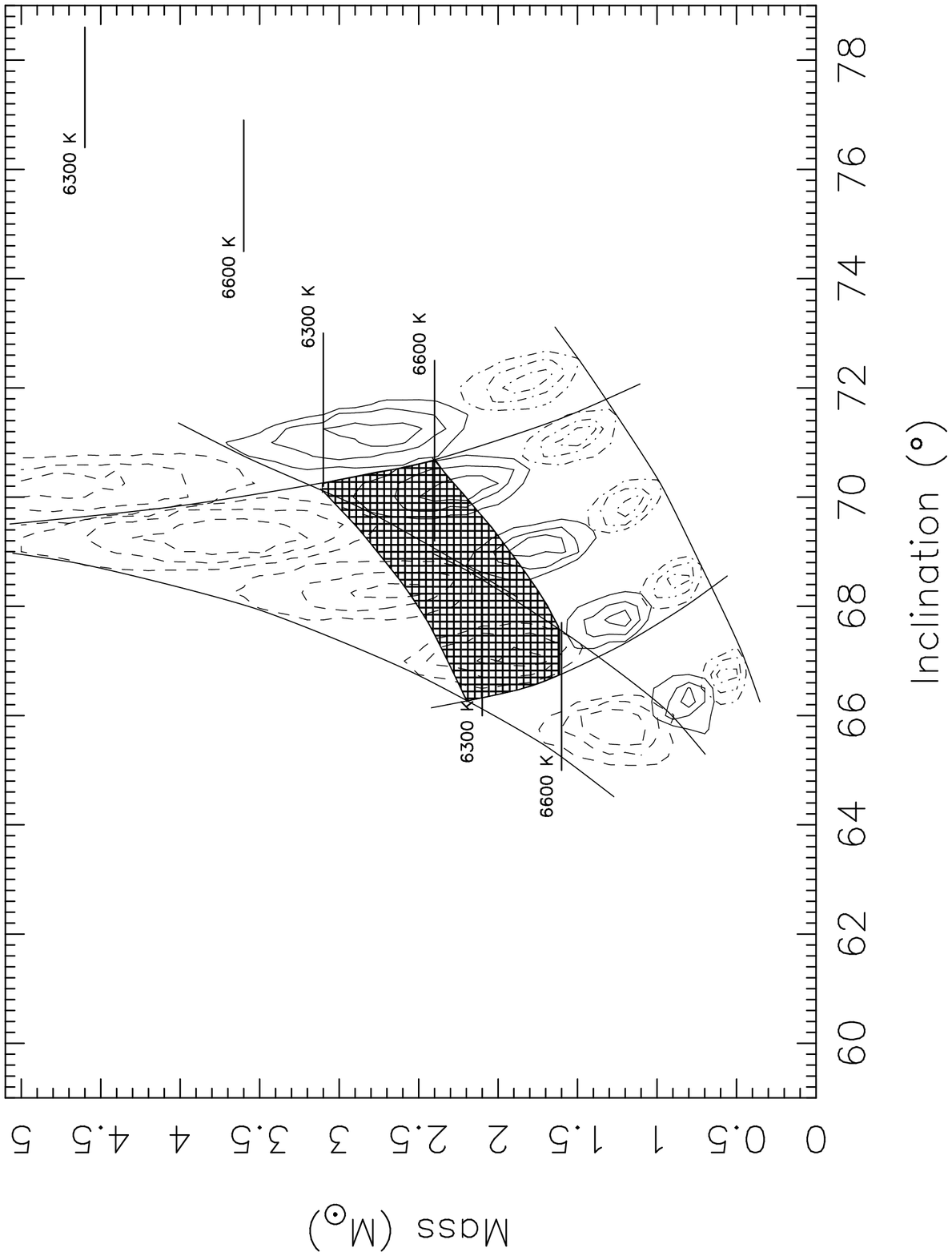,angle=-90,width=9cm} 
} 
\hbox{
\psfig{figure=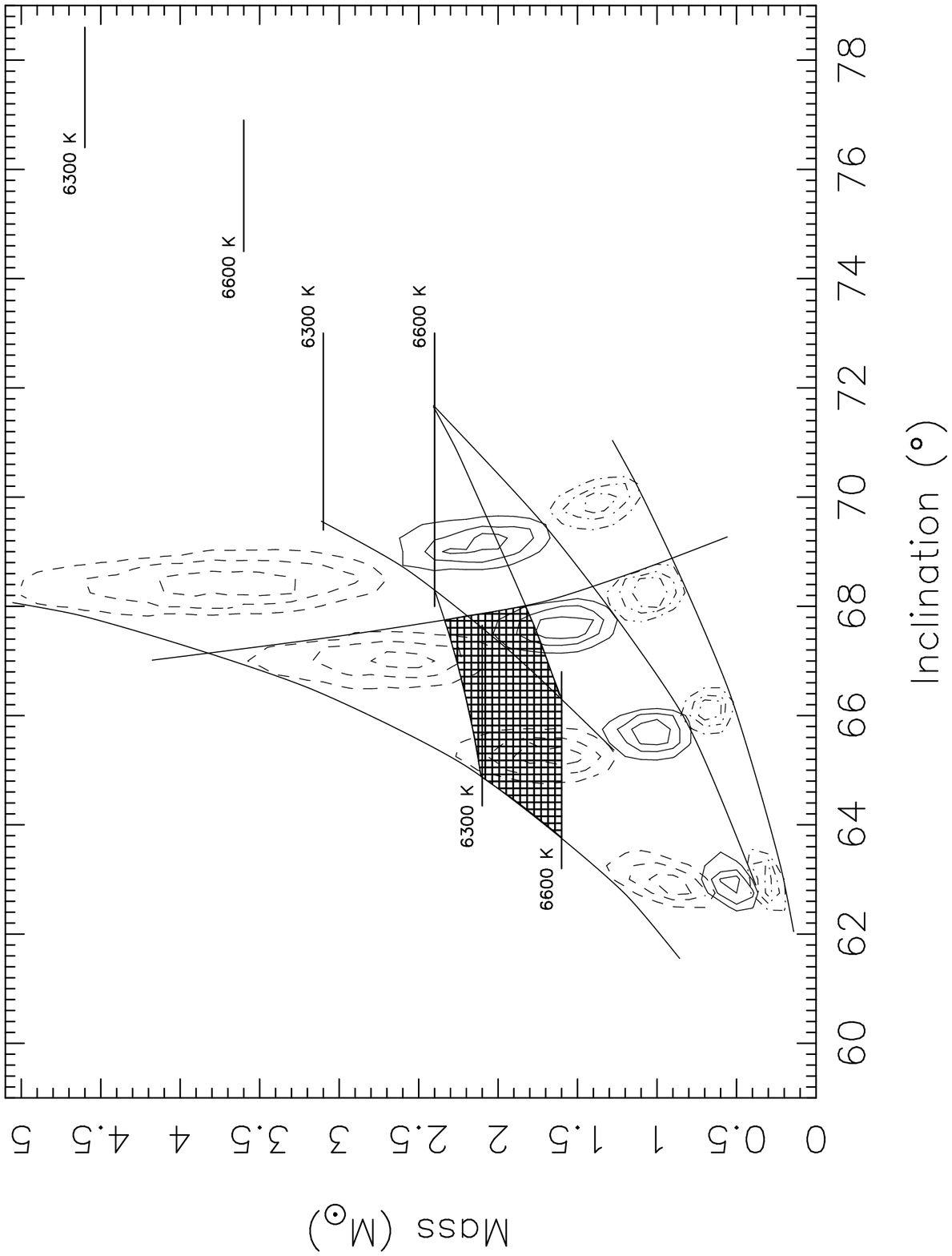,angle=-90,width=9cm}
\psfig{figure=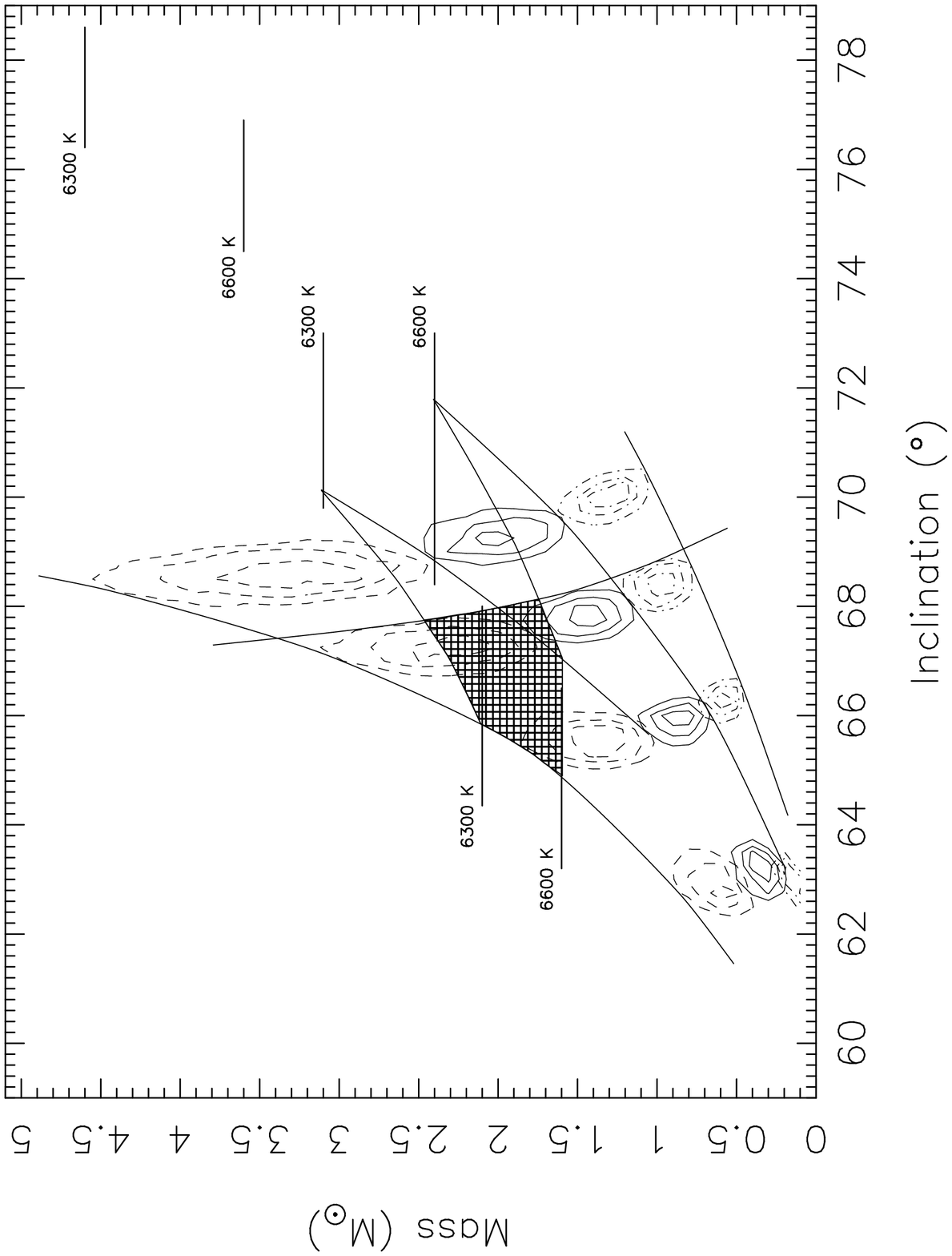,angle=-90,width=9cm} } 
\caption[]{ Total
collection of solutions in the $i$ versus $M_2$ plane for ({\it T$_{\rm
edge}$}, $\gamma$)=(100 K, $2\degr$) [top-left], (1000 K, $2\degr$)
[top-right], (100 K, $10\degr$) [bottom-left], and (1000 K, $10\degr$)
[bottom-right]. The hatched area's are constructed similar as in
Fig.~\ref{fig5}. 
} 
\label{fig6} 
\end{figure*}

\begin{figure*} 
\psfig{figure=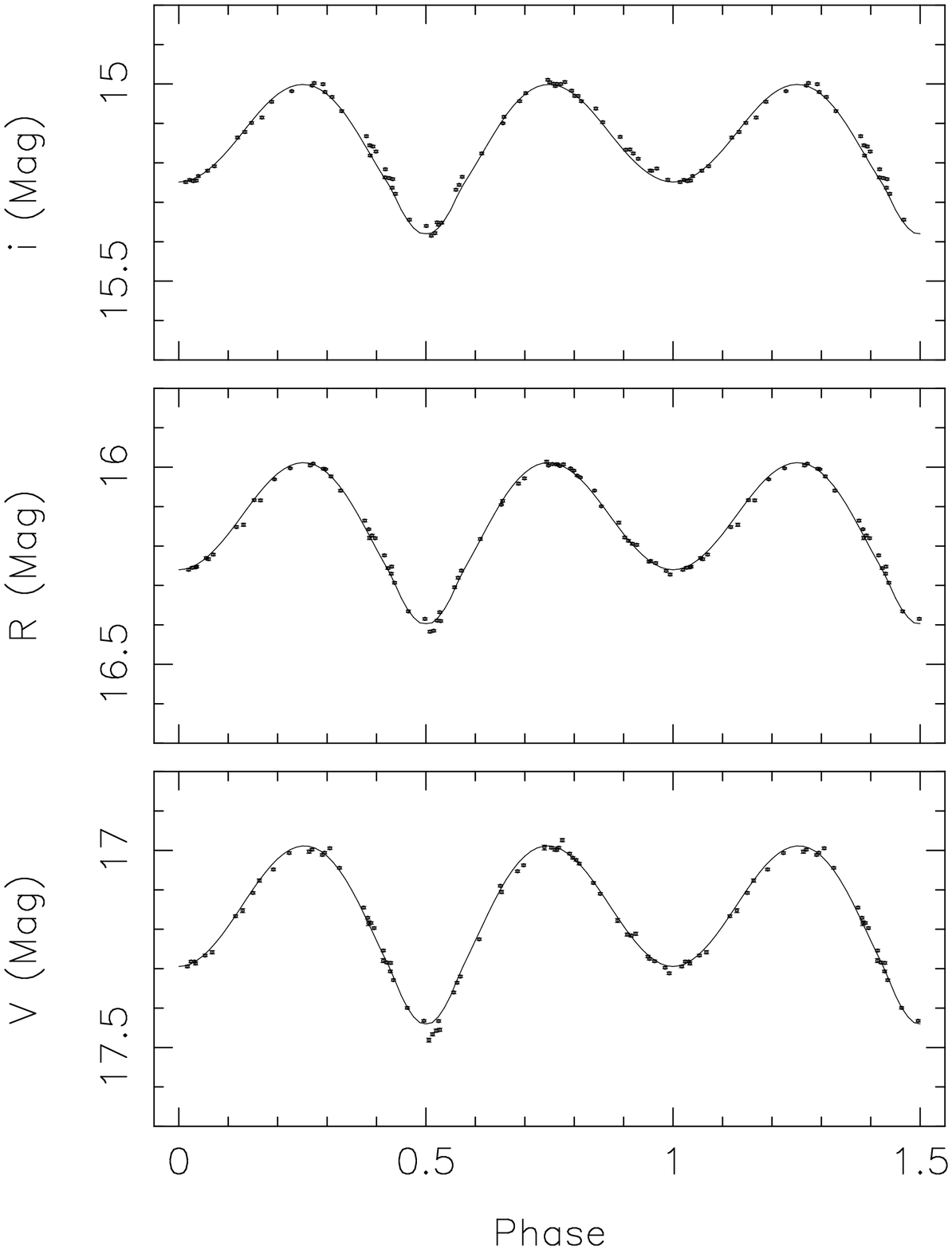,height=22cm} 
\caption[]{ Photometric
data as presented in Fig.~\ref{fig2}, from which six data points have been
removed (see text). The drawn lines represent a theoretical light curve
computed for a binary inclination and secondary mass of (68$\fdg75$, 2.10
M$_\odot$), an intrinsic luminosity of 41 L$_\odot$ and a disk with a 
fractional size of 0.80, a flaring angle of $2\degr$ and a temperature at the 
outer edge of 100 K.
}
\label{fig7} 
\end{figure*}

\begin{figure*} 
\psfig{figure=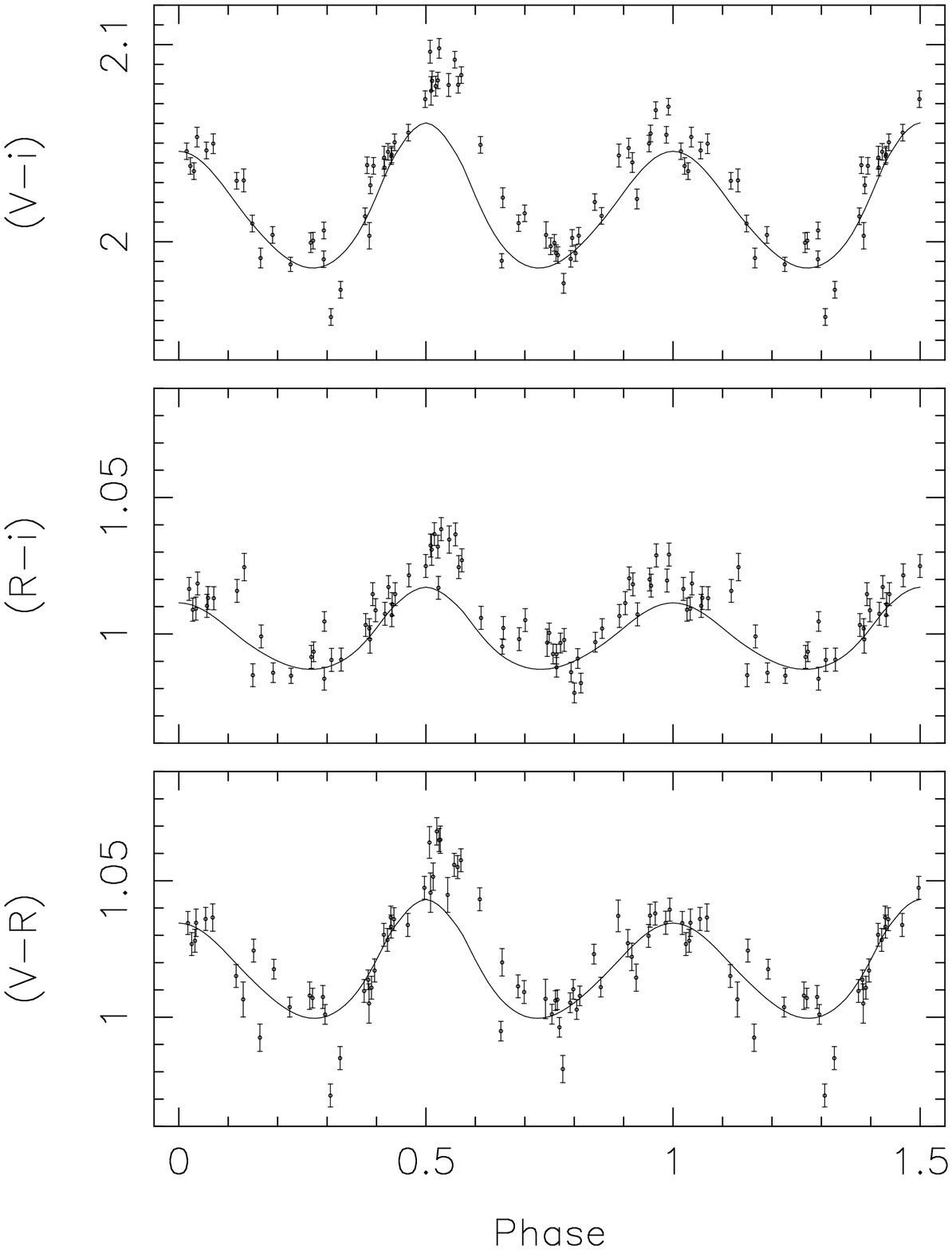,height=22cm} 
\caption[]{
Photometric data as presented in Fig.~\ref{fig3}, from which six data points
have been removed (see text). The drawn lines represent a theoretical color
curve computed for a binary inclination and secondary mass of (68$\fdg75$, 2.10
M$_\odot$), an intrinsic luminosity of 41 L$_\odot$ and a disk with a 
fractional size of 0.80, a flaring angle of $2\degr$ and a temperature at the
outer edge of 100 K.
} 
\label{fig8} 
\end{figure*}

Our modelling of the light curves has led to constraints on the system 
parameters, which have been summarized in Fig.~\ref{fig6}, in the form of 
allowed regions in the $(i, M_2)$ diagram. These regions incorporate solutions 
over the full range in bolometric luminosity of the secondary. From 
Fig.~\ref{fig6} it
appears that there is some dependence of the solutions on the assumed value of
the flaring angle $\gamma$ of the disk, but that it is virtually independent of
its temperature structure. The reason is that the light curve away from phase
0.5 does not allow the disk to contribute substantially to the optical emission
of the system, which sets an upper limit to its outer temperature of 
$\sim1000$~K; 
for these temperatures the main effect of the disk is to eclipse the
secondary. Since for a given value of $i$ the fraction of the secondary that is
eclipsed increases with $\gamma$, we find an anti-correlation between $\gamma$
and $i$ for the allowed solutions (compare the top and bottom panels in 
Fig.~\ref{fig6}). 

\begin{figure*} 
\psfig{figure=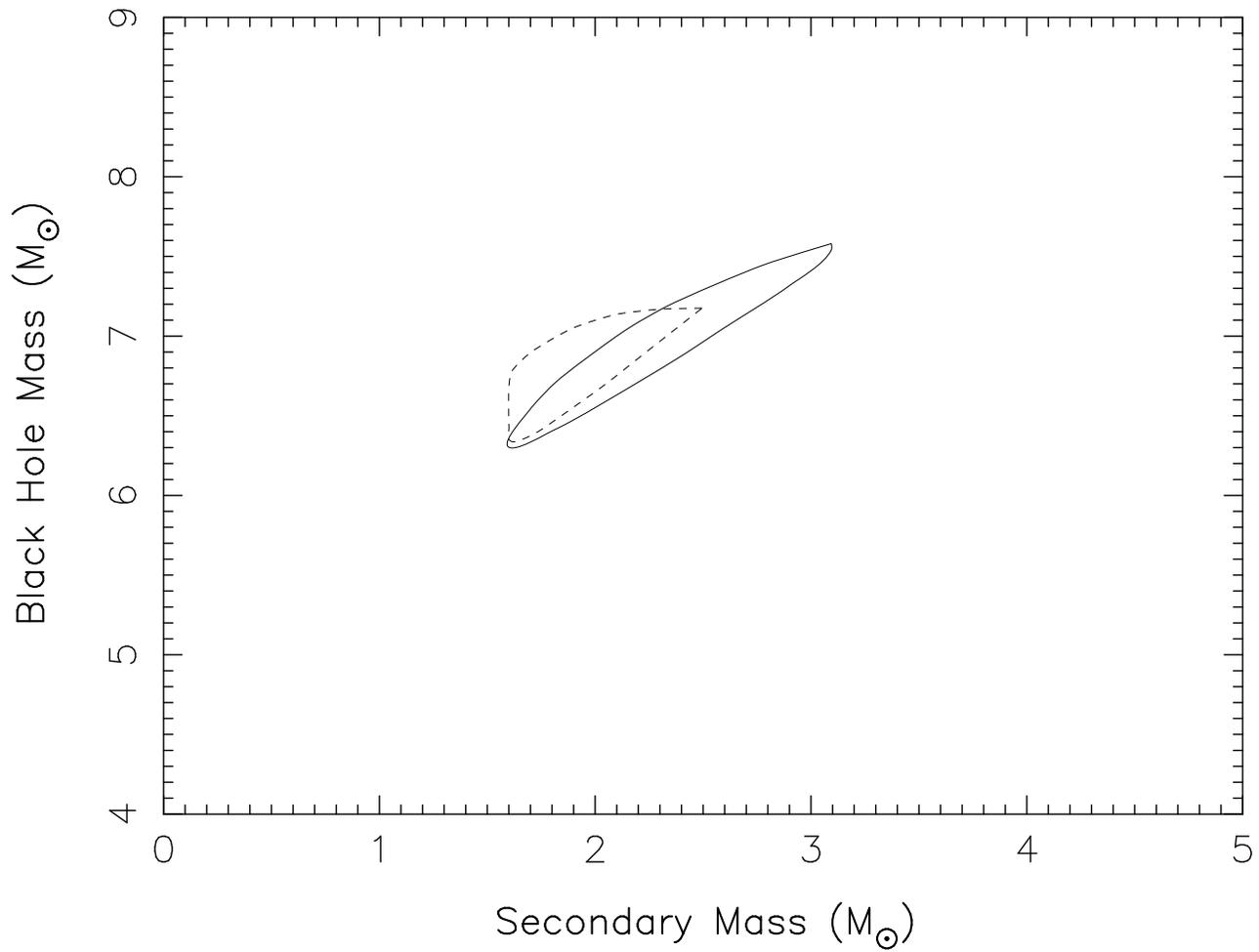,width=17cm,angle=-90} 
\caption[]{
Collection of solutions in the $M_2$ versus $M_1$ plane for a flaring angle of 
the accretion disk of $2\degr$ (solid lines) and $10\degr$ (dashed lines). The 
mass of the secondary star is restricted to 1.6--3.1 M$_{\odot}$, while the 
mass of the compact object lies in the range 6.29--7.60 M$_{\odot}$. 
} 
\label{fig9} 
\end{figure*}

Since the mass function of the system is relatively well determined, a region 
in the $(i, M_2)$ plane allowed by the light curves corresponds to a range of 
allowed values of the mass ratio and the mass of the compact object in \sco. 
For $\gamma=2\degr$ 
we find inclination angles in the range $66\fdg2-70\fdg7$ and
secondary masses between 1.6 and 3.1 M$_{\odot}$. This then implies a mass
ratio, $q$, and mass of the compact object, $M_1$, in the ranges 2.43--3.99 and
6.29--7.60 M$_\odot$, respectively. For $\gamma=10\degr$ the allowed ranges
of $i$ and $M_2$ are $63\fdg7-68\fdg1$, and 1.6--2.45~M$_{\odot}$,
respectively. The corresponding allowed ranges of $q$ and $M_1$ are 2.93--4.20
and 6.35--7.18 M$_\odot$, respectively. In Fig.~\ref{fig9} we present the 
region in the $M_2$ versus $M_1$ plane occupied by the collection of allowed 
solutions, for a flaring angle of the accretion disk of $2\degr$ (solid lines) 
and $10\degr$ (dashed lines). 

\subsection{Theoretical evolutionary tracks} 
%evolution tracks; Hammi
Since the bolometric luminosity and effective temperature of the secondary 
are well constrained, it is possible to compare the position of the secondary 
star of \sco\ in the 
Herzsprung-Russell diagram (HRD) with theoretical evolutionary tracks. 
We have calculated tracks for both single stars and stars in a binary system 
with a recent version of the evolution code developed by Eggleton (1971, 1995). 
The equation of state and the opacity tables have recently been discussed 
by Pols \ea\ (1995). The mass loss in the binary calculations was assumed to be 
conservative and in all calculations we used a metallicity of $Z=0.02$. 

\begin{figure*} 
\psfig{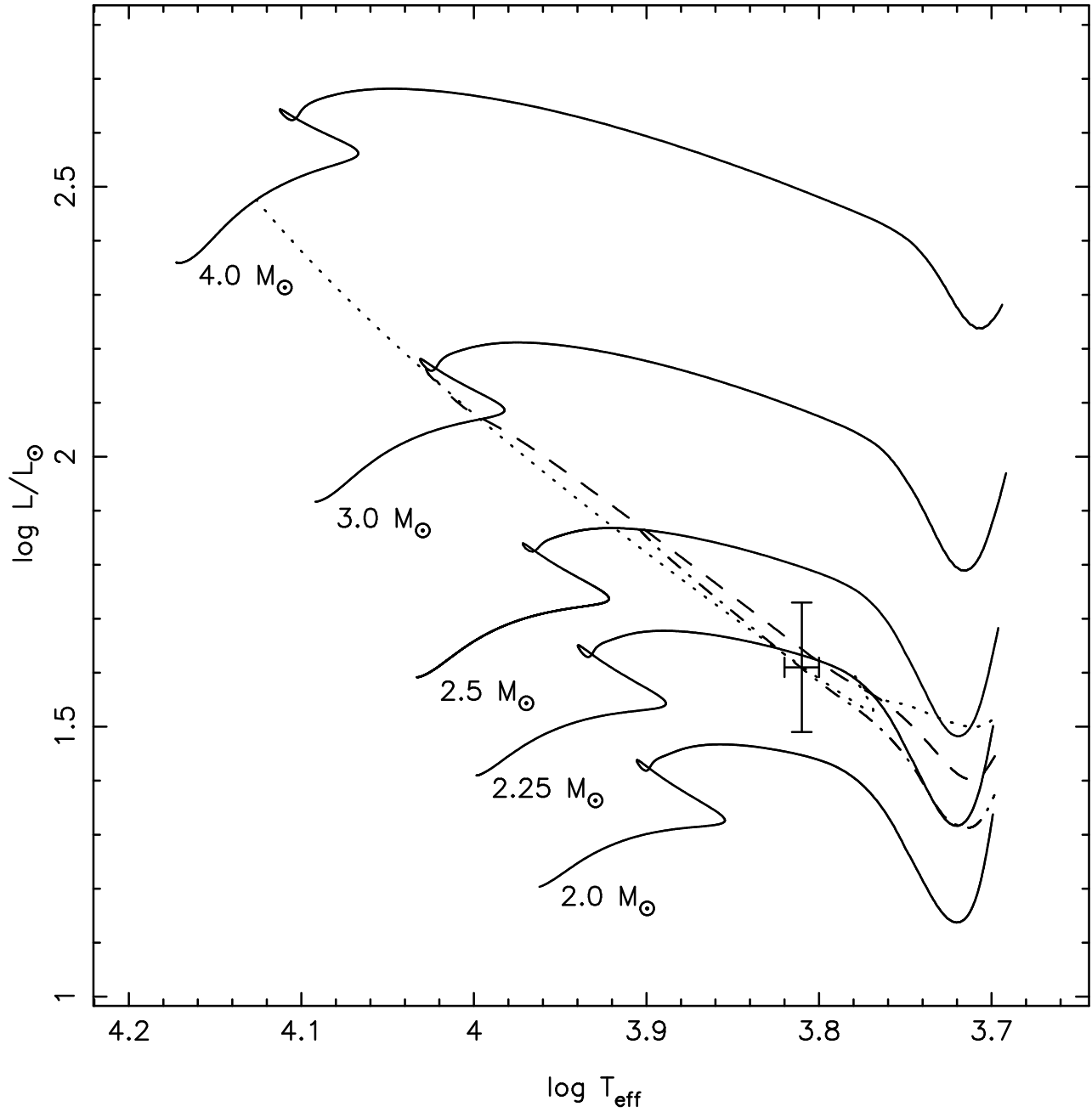} 
\caption[]{
Evolutionary tracks of single stars with mass 2.0, 2.25, 2.5, 3.0 and 4.0 
M$_{\odot}$ (solid lines) and for stars in binary system which lose mass to 
their companion (dashed/dotted curves). 
The total system mass of the binary system is in all cases 9.2 M$_{\odot}$.
} 
\label{fig10} 
\end{figure*}

In Fig.~\ref{fig10} we present a HRD with theoretical evolutionary tracks of 
single stars with masses 2.0, 2.25, 2.5, 3.0, and 4.0 M$_{\odot}$ (solid lines).
Based on this figure, we position the secondary of \sco\ on the track for a 
single star with a mass of $\sim2.2$ M$_{\odot}$. Assuming a mass of 7 M$_{\odot}$ 
for the black hole primary, we obtain an orbital period of about 2.6 days 
(by adopting a radius for the Roche lobe filling secondary from the evolution 
track at the center of the error box [see Fig.~\ref{fig10}], applying Kepler's 
third law and the relation between orbital separation and effective Roche 
radius of a star). This value is in good agreement with the observed period of 
2.62 days. However, this is not the only evolutionary track that fits the 
observations. 

Since the secondary is losing mass, it must have been evolved from an initially 
more massive star. Therefore, we computed several evolution tracks of the 
secondary star in a black-hole binary system, for which we adopted in all 
cases a total system mass of 9.2 M$_{\odot}$. The effect of a different value 
for the total system mass (8--11 M$_{\odot}$) does not change the outcome of 
our results significantly. 
It follows from Fig.~\ref{fig10} 
that a secondary with an initial mass as large as 4.0 M$_{\odot}$ can also 
evolve towards the current HRD position of the secondary of \sco. When the 
tracks of these stars intersect the error box bounded by 
$1.49 \leq \log L_{\rm opt}/{\rm L}_{\odot} \leq 1.73$ and 
$3.80 \leq \log T_{\rm eff} \leq 3.82$, 
the orbital period and mass are about equal to the case of an evolved single 
star. This is not surprising as the secondary is less massive than the 
accreting star, fills its Roche lobe and starts losing mass on the 
main-sequence. The mass loss then occurs on a nuclear time scale, as is the 
case for a secondary with an initial mass of 4 M$_{\odot}$. For the 
secondaries with lower initial masses, the mass loss starts further on the 
main-sequence or even at the moment when the secondary is already crossing the 
Hertzsprung-gap. Since the radius is expanding at a much higher rate, the time 
scale for mass loss is now smaller than the nuclear time scale, but still a 
factor of 10 larger than the thermal time scale. For an initial mass of the 
secondary $\la 4.0$ M$_{\odot}$, the present mass and orbital period for all 
tracks of the mass-losing star that cross the center of the error box shown 
in Fig.~\ref{fig10} are between 2.0--2.3 M$_{\odot}$ and 2.6--3.1 days, 
respectively. Therefore, we conclude that the binary nature of \sco\ is not 
very important for the determination of the present mass of the secondary 
based on evolutionary tracks. 

From standard evolutionary calculations for single stars, we find that the 
tracks of stars with a mass in the range 2.1--2.4 M$_{\odot}$ are consistent 
with the current position of the secondary of \sco\ in the HRD 
(see Fig.~\ref{fig10}). This method to determine the mass of the secondary star
is entirely independent of the constraints derived in Sect.~3.3 from
modelling the light curves of \sco. We emphasize that
the mass range derived from evolutionary arguments is consistent with
that obtained from modelling the light curves.

\subsection{Comparison with Orosz \& Bailyn (1997)} 
Our results are consistent with those obtained by Orosz \& Bailyn (1997). 
These authors describe the quiescence light curve of \sco\ (obtained during 
the same period as our data) in terms of an ellipsoidal model and an 
accretion disk. The main differences between their analysis and the one 
discussed here are the value of the power-law exponent of the radial 
temperature distribution on the accretion disk (fixed to --0.75 in this work, 
free parameter in the analysis by Orosz \& Bailyn 1997), and the use of the 
temperature at the pole of the secondary star as input parameter (fixed to 6500 
K, Orosz \& Bailyn 1997), in contrast to a range of intrinsic luminosities 
(31, 41, and 54 L$_\odot$, this work) as input to our model.

The best fitting $(i, q)$ values of $69\fdg50 \pm 0\fdg08$ and $2.99\pm0.08$ 
($1\sigma$ statistical errors) found by Orosz \& Bailyn (1997), combined with 
the mass function imply a secondary and black hole mass of $2.34\pm0.12$ and 
$7.02\pm0.22$ M$_\odot$, consistent with our results. Also, the values for 
the fractional disk size ($0.747 \pm 0.010$), and angular thickness of 
the disk ($2\fdg23 \pm 0\fdg18$) derived by Orosz \& Bailyn (1997) are 
consistent with the results discussed here for the case of a thin ($2\degr$) 
accretion disk. However, our analysis shows that models with an accretion disk 
of $10\degr$ angular thickness are also able to describe the data well. 
The contribution of the accretion disk to the total luminosity is 
governed by its fractional size and mean temperature, and therefore, by the 
temperature at the outer edge of the disk and the radial temperature 
distribution on the disk. Orosz \& Bailyn (1997) derived an outer edge 
temperature of $4317 \pm 75$ K, significantly larger than the value we derived 
from our analysis. This probably reflects the different value of the 
power law exponent between both analyses; in our analysis the exponent was kept 
fixed to --0.75 (Pringle 1981), while Orosz \& Bailyn (1997) treated the 
exponent as a free parameter and obtained a value of $-0.12 \pm 0.01$. 
However, it should be noted that in both analyses the contribution of the 
disk to the total luminosity of the system is less than $5\%$, consistent with 
the value derived from spectroscopy during 1996 February (Orosz \& Bailyn 1997).

The errors quoted by Orosz \& Bailyn (1997) are $1\sigma$ internal statistical 
errors, derived from fitting the light curves of \sco. 
Note however, that in our analysis we have included errors caused by 
systematic uncertainties in the properties of the system, e.g. in the 
interstellar absorption its light suffers. Therefore, we allow for a range of 
intrinsic luminosities of \sco\ (31--54 L$_\odot$). We also allow a range of 
values for the parameters describing the size and temperature of the 
accretion disk ($0.7 \leq \alpha \leq 0.9, 2\degr \leq \gamma \leq 10\degr$, 
and $100 \leq T_{\rm eff} \leq 1000$ K), and adopt the $3\sigma$ confidence 
contours (see Fig.~\ref{fig6}). Therefore, the limits to the system parameters 
we have derived in this study are somewhat larger compared to those obtained by 
Orosz \& Bailyn (1997).

\subsection{Systematic effects} 
In our modelling of the light curve of \sco\ we have made a number of implicit
assumptions, e.g., with respect to the geometry of the disk, and such
assumptions may affect the results of our analysis systematically. There is one
assumption whose systematic effect is particularly strong, i.e., that the
non-uniform temperature distribution across the surface of the secondary can be
described by a standard gravity darkening coefficient $\beta=0.25$. This
value applies in a situation where the outer layers of a star are in radiative 
equilibrium. For stars with convective outer layers  Lucy (1967) derived a
gravity darkening coefficient of 0.08. At its spectral type (F3-F6) the
secondary of \sco\ is near the boundary in the Hertzsprung-Russell diagram
separating hot stars with radiative envelopes and cool stars with convective
envelopes. Therefore, it may contain a shallow convective layer. We have
investigated the applicability of standard gravity darkening by computing light
curves for $\beta$ in the range 0.08 to 0.25 (in steps of 0.01). Decreasing
$\beta$ leads to a decrease in the brightness contrast over the secondary's
surface, and therefore to a decrease of the amplitude of the light curve. Since
the amplitude increases with $i$ the strongest limits on $\beta$ are obtained
for $i=90 \degr$. For an assumed $M_2=2.2$~M$_{\odot}$ and 
$L_{\rm opt}=41$~L$_{\odot}$ 
(near the center of our solution space) we derive a lower limit
$\beta \geq 0.18$ on the gravity darkening coefficient. This suggests that the
assumption $\beta=0.25$ is justified. 

\section{Conclusion}  
We conclude that during our March 1996 observations \sco\ was in a state of 
true X-ray quiescence. We derive a refined ephemeris for inferior
conjunction of the secondary star with respect to the compact star of HJD
$244\,9838.4198(52) + 2.62168(14) \times {\rm N}$. We calculated theoretical
light- and color curves, which constrain the binary inclination and mass of the
secondary star to lie in the range $63\fdg7-70\fdg7$ and 1.60--3.10 M$_\odot$,
respectively. The implied mass of the black-hole is in the range 6.29--7.60
M$_\odot$. We find that an accretion disk is required to model the light- and
color curve of \sco. The disk does not contribute significantly to the
luminosity, and therefore, acts as body of obscuration only. The mass range we 
obtained for the secondary star is supported by the results of stellar 
evolution calculations. 

\begin{acknowledgements}
The authors thank F. Kemper for her 
help in obtaining the observations. FvdH acknowledges support by the
Netherlands Foundation for Research in Astronomy with financial aid from the
Netherlands Organisation for Scientific Research (NWO) under contract number
782-376-011. 
\end{acknowledgements}

%%============================================================= 
%	References
%%============================================================= 

% 
%%============================================================= 
% 	Figures
%%============================================================= 

\end{document}